\documentclass[manuscript,screen,nonacm]{acmart}

\begin{document}

\title{Rationality in current era - A recent survey}

\author{Dibakar Das}
\orcid{0000-0002-6731-8975}
\email{dibakard@acm.org}
\affiliation{%
  \institution{IIIT Bangalore}
  \state{Karnataka}
  \country{India}
  \postcode{560100}
}

\begin{abstract}

  Rationality has been an intriguing topic for several decades. Even the scope of definition of rationality across different subjects varies. Several theories (e.g., game theory) initially evolved on the basis that agents (e.g., humans) are perfectly rational. One interpretation of perfect rationality is that agents always make the optimal decision which maximizes their expected utilities. However, subsequently this assumption was relaxed to include bounded rationality where agents have limitations in terms of computing resources and biases which prevents them to take the optimal decision. However, with recent advances in (quantum) computing, artificial intelligence (AI), science and technology etc., has led to the thought that perhaps the concept of rationality would be augmented with machine intelligence which will enable agents to take decision optimally with higher regularity. However, there are divergent views on this topic. The paper attempts to put forward a recent survey (last five years) of research on these divergent views.
  These views may be grouped into three schools of thoughts. The first school is the one which is sceptical of progress of AI and believes that human intelligence will always supersede machine intelligence. The second school of thought thinks that advent of AI and advances in computing will help in better understanding of bounded rationality. Third school of thought believes that bounds of bounded rationality will be extended by advances in AI and various other fields. This survey hopes to provide a starting point for further research.
\end{abstract}


\maketitle

\section{Introduction}
Rationality has been an intriguing topic for a long time. Even, the meaning and definition of rationality differs across various subjects, such as, cognition, philosophy, psychology, anthropology, sociology, political science, economics, and mathematical and computational theories \cite{cite_scope_of_rationality}\cite{cite_different_definitions_of_rationality}. In certain contexts, agents acting in self interest have been termed as rational behaviour. Mathematically, maximizing each agent's expected utility (global optimum) always is also defined as (perfect) rational behaviour \cite{cite_rationality_as_maximize_utility}. Some scholars disagree with the mathematical formulation of rationality and consider the topic in much broader sense of environment, culture, etc. \cite{cite_book_rethinking_rationality}.

However, the view of perfect rationality have been contradicted with the notion of bounded rationality for humans \cite{cite_handbook_of_bounded_rationality}.
In bounded rationality, agents (e.g., humans) tend to take decisions which are not necessarily the optimal always due to inherent biases and limited resources to compute. Mathematically, it could be interpreted as sub-optimal decision making (a local optimum). Many scholars suggest that irrational behaviours which could be sub-optimal decisions mathematically should not be interpreted as bounded rationality and should be seen from a broader scope of different subjects \cite{cite_book_rethinking_rationality}\cite{cite_rationality_as_maximize_utility_not_very_useful}. Comprehensive study of bounded rationality and its implications are presented in \cite{cite_handbook_of_bounded_rationality}. Interrelation between bounded rationality and perfect rationality is studied in \cite{cite_rationality_bounded_rationality_same} where authors argue that context of (perfect) rationality is itself dependent on bounded rationality.

With the recent advances in the field of computation, artificial intelligence (AI), science and technology, etc., there has been a growing thought on humans, being assisted with machine intelligence, achieving optimal decision and moving towards higher levels of rationality \cite{cite_ai_will_enhance_the_bounds_bounded_rationality_marwala}. Several question arise in this context \cite{cite_21th_century_singularity}. Will humans be above machines? Will machines overtake humans? Will it make humans perfectly rational? Will it improve the bounds of bounded rationality? Will the advances in various fields lead to better understanding of rationality? Several thoughts exist in the literature. The paper tries to capture some of the recent works in the last five years on this field of how human rationality will be impacted by advances in different fields in a broader context without going into specific use cases which are limited in scope.

For many practical purposes, rationality can be interpreted as taking the optimal decision though it may not be in the broadest sense of the term in different subjects. For example, taking the optimal decision by agents (assisted by machines) to a problem (always) cannot be beyond the definition of rationality in its broadest sense. Also, a sub-optimal decision to a problem taken by an agent due to lack of computational resources or biases cannot be outside the realm of bounded rationality in its widest scope. Thus, the terms rationality and bounded rationality may have wider implications in different subjects, however, mathematically and computationally taking a decision, whether optimal or sub-optimal, does fall with the scope of the two terms.

The term computational rationality has received quite a bit of attention in the recent years. This topic essentially is a convergence of cognitive, AI and computational neuroscience with the intention to take decisions which maximizes expected utility \cite{cite_computational_rationality_review}\cite{cite_computational_rationality_framework_information_processing}. Thus, machines will complement bounded rationality of humans to reach the optimal solution.

Application of quantum theory to rationality has been an area of great interest in recent times.
Considering two-stage gambling and prisoner's dilemma games, \cite{cite_quantum_probability_gambling_prisoners_dilemma} shows how quantum probability model can explain the violations in sure-thing principle of decision theory.
Quantum model for Ellsberg Paradox have also been proposed in \cite{cite_quantum_model_for_ellsberg_paradox}.
Apart from using quantum theory, \cite{cite_quantum_and_phenomenal_research_for_cognition_and_bounded_rationality}
highlights the importance of phenomenological research to augment the shortcomings of the former. \cite{cite_quantum_models_for_cognitive_information_processing} argues that quantum-like models very well match the specifics of cognitive information processing.

One of the earliest work on comparison of bounded rational agents and unbounded rationality is \cite{cite_repeated_game_bounded_rational_and_unbounded_rational}. This paper explores the scenario of two types of agents in a two-player repetitive game and finds that the unbounded rational player has a dominant strategy. In a repeated zero-sum game, the unbounded rational player has a strategy to bring down the payoff of the bounded rational player.
With recent advances in AI and computing for the past one and a half decade, if agents are bounded rational and they are aided by machines to help obtain the optimal decision most of the time, then bounds of bounded rationality are likely to be extended \cite{cite_ai_will_enhance_the_bounds_bounded_rationality_marwala}.

As expected of a new thought process, there are divergent interesting views on this topic of rationality augmented machine intelligence. The approaches are so varied that classifying them is a major challenge. In an attempt to broadly classify recent works in this sphere, three categories of thoughts seems to emerge. One school of thought is that bounded rationality will remain despite advances in computing \cite{cite_cognitive_heuristics_in_ai_will_exist}. One group of research works believe that advances in computing and other related fields will enhance the understanding of bounded rationality \cite{cite_information_process_limitation_human_behaviour}. The third group believes that the advances in computing, AI, etc., can actually enhance to bounds of bounded rationality towards perfect rationality \cite{cite_ai_will_enhance_the_bounds_bounded_rationality_marwala}.

This paper surveys some of the recent works in last five years on interesting and divergent views on the implication of major advances in different fields on rationality.

\subsection{Living with bounded rationality}
Although, there is enough enthusiasm on application of recent advances in computing, AI, etc., to improve human decision making, some research works are sceptical about these progresses in taking humans closer to perfectly rationality. Some of the related views are surveyed below.

There are some thoughts that human will continue to remain bounded rational and applications of approaches from bounded rationality will continue to be relevant. \cite{cite_cognitive_heuristics_in_ai_will_exist} opines that heuristics applied due to bounded rationality has played a central role in AI research. The author argues that cognitive-based heuristics will still be relevant for applications which are easily solved by humans and hard for machines. This shows the continued relevance of cognitive heuristics in the realm of automated decision making. \cite{cite_chess_player_deviate_bounded_rationality} show that even deviation from bounded rationality (let alone perfect rationality) does not necessarily leads to losses. Players tend to deviate from bounded rationality benchmark of a chess engine due to various circumstantial factors such as fatigue, complexity, pressure from being in a better or worse position and time taken to decide. Results show that faster decisions leads to more deviations. However, deviations from bounded rationality often gives better returns regularly probably due to experience and intuition of the players.
Author in \cite{cite_ai_cant_outsmart_knowledge_work} argues that AI cannot replace human in complex work environments which involves relationships, societal aspects and contexts.

\subsection{Augmenting rationality}
There is an overwhelming feeling among research works that recent advances in computing, AI, science and technology, etc., will bring a change in human decision making with augmented machine intelligence. However, there is divergence of views on the impact of augmented machine intelligence on human society. This section surveys some of the most interesting research works in this direction.

There have been several works on how computational and AI advancement can aid humans where needed. To overcome bounded rationality of humans, virtual world digital nudging has been studied to push agents towards more rational decisions. However, this nudging requires ethical consideration in the way they work. \cite{cite_ethics_digital_nudges_for_rationality} presents guidelines for designing such ethical digital nudges.
An argument has been put forward in \cite{cite_ai_limits_human_bounded_rational_machines_more_rational} which claims that although humans are bounded rational, they do not pass on their limitations to the machines. Thus, machines can be made more rational, though no rigourous proof has been attempted. Author also claims that technology can improve bounds of rationality (flexibly-bounded rationality) but always lower than perfect rationality. \cite{cite_computational_rationality_rl_motor_control_rate_distortion_theory} analyzes the notion of computational rationality (maximization of utilities subject to information processing constraints). It applies rate-distortion theory to model computationally rational agents. This thesis also extends the model to reinforcement learning and motor control. %
In this thesis work \cite{cite_computational_relevance_use_of_information_without_scrutiny}, author brings forward a very interesting aspect of using information with scrutiny in decisions under bounded rationality which is named as \emph{bounded relevance}. The author argues that using readily available information from various sources in today's world without proper scrutiny can lead to disastrous decision making.
Interaction between humans and intelligent software agents is studied in \cite{cite_interaction_humans_intelligence_sw_agents}. Assuming bounded rational agents and applying concepts from artificial intelligence, behavioural economics, control theory, and game theory, agents acquire rewards from actions of the humans. Authors argues as agents learn from humans the former steers humans from optimal decisions, lowering autonomy of humans and forcing them to addictive and compulsive behaviour. They also consider ethical, social and legal implications of such intelligent agents and opines that these can exploit and reinforce human weakness.
\cite{cite_computational_rationality_review} reviews the progress of \emph{computational rationality} of taking actions with highest utility with the aim to understand choice of action under uncertain condition. Author review advances in large scale probabilistic inference, computational cost, precision and timeliness. The interaction between three related fields of computer science, cognitive science, and neuroscience is also analysed. %
Decision making under bounded rationality has been a widely researched area. A data driven decision theory integrating recent trends in big data analytics, machine learning, automated decision making has been explored in \cite{cite_big_data_ml_decision_bounded_to_collaborative_rationality}. This work emphasizes the ardent need to integrate traditional decision making mechanisms with the emerging topics in computing to make better and informed choices and hence, going beyond bounded rationality to \emph{collaborative rationality}. 
In \cite{cite_digital_x_from_bounded_rationality_to_bounded_imagination}, authors argue that digitalization has decreased the challenges of bounded rationality in terms of parameters, such as, time, knowledge and resources. However, authors claim that this phenomenon has led to a new challenge of \emph{bounded imagination} to visualize and conceptualize the implication of digitalization in future.
An information theoretic view of relationship between economics, psychology and artificial intelligence in the context of decision theory has been reviewed in \cite{cite_review_info_theory_ai_psycho_rationality}. Key aspect of this review is to study the relation of expected utility theory to rationality, bayesian decision making and information theory. Author argues that models based on the above relationships are computable with necessary computing resources and the results can be compared to human abilities. Authors conclude the work with the thought that probably several questions remains unanswered for AI enabled agents to take decision as good as humans.
An information theoretic approach to decision making under Smithian competition paradigm has been investigated in \cite{cite_info_theoretic_how_agents_behave_with_extra_info}. The model captures how bounded rational agents have to acquire information to augment their prior beliefs which is measured with Kullblack–Leibler (KL) divergence between decisions after acquiring new knowledge and prior ones. Agents acquire information easily they tend to go by new insights whereas when information is difficult to acquire they tend to go back to their previous belief. Maximum entropy information is used to decide on least biased decisions under Quantal Response Statistical Equilibrium in Smithian competition framework. Studies show that agents change their decisions when new information is available. The study also shows some indication of evolution of agents beliefs.
In this interesting work \cite{cite_3_views_of_expertise_ai_possibilities}, authors study expertise from three different views, namely, Dreyfus and Dreyfus’s View on Expertise (1986),  Gobet and Chassy’s View on Expert Intuition (2009) and Montero and Evans view on expertise (2011) from philosophical implications for rationality,
knowledge, intuition and education. They also consider the three views from context of AI. Dreyfus and Dreyfus’s View on Expertise seems to suggest that AI systems can perform as best as expert humans and not better. Gobet and Chassy’s (2009) View on Expert Intuition suggests that intuitive knowledge cannot be built into intelligent systems. However, they agree that there is possibility of declarative and procedural knowledge getting incorporated into AI systems. For Montero and Evans view on expertise, expert knowledge can be coded into expert systems and can be communicated to bounded rational humans.
An agent based model to study how intelligence in rational agents evolve with knowledge from agents' interactions is explored in \cite{cite_agents_knoledge_centralization_and_divide}. Agents acquire knowledge by a selection-decision rule. The model studies various aspects, such as, how knowledge evolves in agents, innovation rate of the agents, top innovators among agents and impact of network size. Results reveal that emergence of rational intelligent agents lead to centralization in the network, and average knowledge and knowledge inequality rises exponentially.
Author studies the impact of machines defeating humans in Go game \cite{cite_beyond_human_rationality_to_machine_rational_go_game} and opines that these form of games are machine bounded rational and are different from human bounded rationality. A cellular automata is used to model the Go game as a computational process. Since, cellular automata are undecidable, author concludes that machine bounded rationality has limitations though they tend to surpass humans with bounded rationality in certain situations.
In \cite{cite_ethical_ai_and_intrusion_on_liberal_values}, authors consider the role of data and AI trends and their ethical usages. They argue that regulations applied to liberalism, privacy and autonomy of individuals, and ethics are not compatible with the changes happening due to AI. These data driven AI techniques are infringing on human aspects. They also argue that organizational usage of AI is trying to extend bounded rationality and leading to inequality serving only the shareholder's value. Authors highlight the need of ethical use of AI and proposes a system that has prioritized control over incoming and outgoing data as a way to preserve autonomy and liberal values of individuals.
\cite{cite_artificial_agents_abilities_in_decision_making_framework} develops a framework to test ability of \emph{Individual Evolutionary Learning} agents  in human decision making in a Voluntary Contributions Mechanism experiment. Through this framework authors find that the ability of the agents to mimic human behaviour does not necessary indicate their decision making abilities. This work suggest some way to improve the decision making ability of the agents.
Application of AI in organizational decision making under uncertainty is studied in \cite{cite_survey_ethical_ai_organization_decision_making}. This work bring out the challenges, pre-conditions and possible consequences of use of AI by humans in decision making. Also, the authors recommended use of ethical AI frameworks. Otherwise, the consequences of using AI can actually create more problems then solving things. Hence, educating workforce on responsible use of AI is important.
A study of recent trend in increasing dysfunctional disagreement under information overload is presented in \cite{cite_info_overload_bayesian_inference_do_not_work_bn_qpt_works}. Bayesian rationality in agents does not seem to work under information overload. Authors provide approaches of using bayesian networks or formation of knowledge partitions (quantum probability theory) to deal with information overload.
In this work \cite{cite_ai_double_edged_sword_reinforcement_effect_on_individual}, authors conclude that AI can augment human intelligence and support individual's and organization's development. However, over dependence of AI in organizations can lead to reinforcement effect which essential means that strengthening of certain future behaviour given antecedent stimulus.
A framework to show much AI agents can replace humans in organization is described in \cite{cite_transform_human_to_ai_agents_in_org}. The paper also provides a way to move forward in this digital transformation. Authors review concepts from Carnegie School and behavioral theory, and discuss the need for innovation management in AI based systems.
In this work \cite{cite_wisdom_service_cognitive_negotiators_btwin_humans_machines}, author introduces an interesting concept of wisdom service systems which introduces cognitive mediators which can interact with machines and humans. These mediators can adapt to varying situations and also suggest recommendations.
Applying computational rationality author presented a framework of building artificial moral agents integrating philosophical and scientific aspects to advance machine ethics \cite{cite_artificial_moral_agents_using_computational_rationality}.
Author of \cite{cite_egoism_from_natural_selection_to_optimal}  argues that egoistic behaviours from natural selection theory can lead to higher degree of rationality.
In this work \cite{cite_regularization_of_computable_functions_for_bounded_rationality}, authors consider building solutions to deal with bounded rationality by regularizing solutions of computable functions from specific data and cognitive biases.
\cite{cite_use_of_foresight_to_mitigate_bounded_rationality} demonstrates how foresight can be used to augment bounded rationality.
A collaborative online learning framework among bounded rational agents to make a optimal decision beyond the capabilities of individual agents is presented in \cite{cite_collaborative_online_learning_bounded_rational_agents}.

The above research offers interesting general ideas and scope of how augmentation of data information, computation and AI can have varied impacts on human rationality with the possibility of extending the bounds to higher levels.

\subsection{Understanding bounded rationality in a better way}
Bounded rationality, its interpretation and its implications have not been understood completely though the concept was proposed decades ago. Various explanations have been put forward over time \cite{cite_handbook_of_bounded_rationality} and the process continues till date. With the advent of new concepts in computing, AI, science and technology several, etc., new approaches have been proposed recently to understand the same. Some of  interesting proposals in this topic are surveyed below.

In \cite{cite_marl_gr2_game_theory}, authors present hierarchical levels of agents with varying levels of rationalities in a recursive reasoning framework to recognize bounded rationality of its agents, leading to their sub-optimal behaviors. They also prove existence of perfect bayesian equilibrium. The paper also proposes a practical actor-critic model within the reasoning framework and demonstrates stationary point convergence with Lyapunov analysis. Authors claim that the model performs better than the previous models of agents in normal form games.
Authors of \cite{cite_self_modify_bounded_rational_agents_low_returns} analyze the impact of bounded rational agents which  modify themselves with the intention to increase their utility. Although, it was shown in \cite{cite_self_modify_fully_rational_agents_no_harm} that self modification do not harm perfectly rational agents, \cite{cite_self_modify_bounded_rational_agents_low_returns} shows that this result does not hold good for agents with bounded rationality. Considering four reasons for bounded rationality, inability to choose action optimally, unaligned to human values, inaccurate perception of environment and use of incorrect temporal discounting factor, authors of  \cite{cite_self_modify_bounded_rational_agents_low_returns} show that self modification can lead to exponential decrease in performance.
For competitive multi agent systems, \cite{cite_multi_agent_explore_exploit_q_learn_converges_any_number_of_agents} considers tradeoff between exploration and exploitation using
smooth Q-learning which considers both game rewards and exploration costs.
Authors show that learning converges to the unique quantal-response
equilibrium (QRE) for weighted zero-sum polymatrix games involving heterogeneous players with positive exploration rates under bounded rationality for any number of agents and not requiring any parameter fine-tuning.
Authors of \cite{cite_is_bounded_rationality_bounded_by_unbounded_or_behavioural} tries to analyze whether bounded rationality is upper bounded to unbounded rationality. They come to the conclusion that human beings and organisations tend to behave adaptively based on complex and changing environments, rather than any prescriptive models of rationality and behavioural economics.
Contrary to classical game theory, a bounded risk-sensitive Markov game has been used for forward design policy and inverse reverse learning in multi-agents systems \cite{cite_bounded_risk_sensitive_markov_game}. Using concept from prospect theory, the paper assumes bounded intelligence and maximization of risk utilities. Results show that agents show both risk-averse and risk-seeking behaviours. For inverse reward learning, the proposed model outperforms risk-neutral approaches in terms of acquiring accurate reward values, and intelligence levels and risk parameters of agents.
Influence of knowledge dynamics in decision making has been explored in \cite{cite_multi_field_entropic_knowledge_dynamics} using  multi-field theory of organizational knowledge and the entropic knowledge dynamics model. Authors compare this approach with the bounded rationality and intuition dimensions which lead to integrated multi-factor conceptual and structural model of decision making. Authors show relevance of entropic knowledge dynamics with detailed surveys.
In \cite{cite_information_process_limitation_human_behaviour}, author shows that Hick-Hyman law and the Power Law of Learning result from inherent limitation of information processing in humans applying stochastic techniques. Considering cognition as a information transmission bounded process,  \cite{cite_information_process_limitation_human_behaviour} shows that it can lead to consolidated explanation of human behaviour under bounded rationality. 
Author in \cite{cite_bounded_rationality_using_rl} uses RL as a model to define bounded rationality. Using this model, rational agents form concept due to resource constraints. Further, this work tries to define a set of concepts for rational agents with limited resources.
Authors introduce a critical-induced bounded rationality (CIBR) in \cite{cite_critical_induced_bounded_rationality_quantum_probability}. They argue quantum probability seems to better explain  certain functions of the human brain and cognition as compared to bounded rationality defined using classical probability. They study decision making under bounded rationality in the same way as quantum probability to generate choices and arrive at CIBR. In certain scenarios, CIBR seems to perform better than quantum probability. 
In this work \cite{cite_abstraction_hierachical_elipse_roundedness}, authors consider stereotypic actions as a way to minimize information processing and  reduce variance in utility under bounded rationality constraints. Considering the estimation of roundedness of an ellipse, agents could decide on three different levels of precision corresponding to their three hierarchical abstraction levels. However, varying the difficulty in identifying the objects and the response times to estimate the roundedness agents adapt prediction of the abstraction level based on available processing resources. This adaptive behaviour is compared with the limits of bounded rational decision making and identify any inefficiencies.
\cite{cite_solve_tasks_crowdsourced_agents_rationality_bias} investigates the connection between three different task structures, i.e., local tasks, small-world tasks and random tasks, and bounded rationality of individuals on performances in crowdsourced solutions. Bounded rationality levels is used to distinguish industry types,e.g., emerging and traditional. Bounded rationality bias is used to characterize solvers as expert, professional and ordinary users. An agent based model along with NK fitness landscape and TCPE (Task-Crowd-Process-Evaluation) is used to depict crowdsourcing process. Results show when tasks from the three categories are equally complex, random tasks are harder to solve than local tasks. In emerging businesses (bounded rationality level low), local tasks do better than others for any type of crowd solvers. In traditional industries (bounded rationality level high) with ordinary crowd solvers, local tasks are performed best followed by small-world and random task. When crowd solvers are expert and professionals, random tasks do best.
A finite retrospective inference as an improvement over Bayesian smoothing is proposed in \cite{cite_finite_retrospective_inference_hmm_for_bounded_rationality} to model bounded rationality. Finite retrospective inference takes into account only a finite past events and current information to update the agent's beliefs under available computation and other resources. Simulation results show that the model significantly increase the accuracy of inference and learning in hidden markov models.
\cite{cite_uncertainty_modeled_as_ecological_mobility} presents approaches to uncertainty and decision making from the perspective of ecological rationality since human risk analysis cannot be explained completely by probability and decision theories. Authors argue that a system wide view of uncertainty can be better explained by organism-environment interaction in the ecological paradigm. Results show application of simple heuristics can model unmeasurable uncertainty. 
A rigorous mathematical theory of bounded rationality is presented in \cite{cite_rigorous_math_model_rationity_phd_thesis_ucb}. The analysis helps explain the cognitive aspects of decision making and judgement, and also ways to improve these activities. 
Authors in \cite{cite_qualitative_behavioural_economics_to_quantitive}, argues that certain decision with  bounded rationality of humans which seems to rely on qualitative analysis in the behavioural economics can actually be explained quantitatively using numerical approaches.
\cite{cite_inverse_decision_modeling_sequential_process} develops a framework for learning the parameters to represent behaviour in sequential decision making. This framework also opens up learning the representation of bounded rationality including biased beliefs and suboptimal decisions. Authors also study cognitive biases in animals and robots.
In the context of interactive data visualization, \cite{cite_rra_cognitive_biases_data_visual} presents a resource-rationality analysis in cognitive biases using Bayesian cognitive models under bounded rationality constraints of time and computational costs.
Human irrationality (e.g., cognitive biases in Kahneman-Tversky tradition)  or bounded rationality (e.g., contextuality) can be modeled using models from quantum mechanics \cite{cite_cognitive_biases_contextuality_in_rationality_with_quantum_cognitive_science}. These aspects of human rationality are studied with cognitive bias and contextuality experiments to form a perspective on \emph{quantum cognitive science}. Authors argue that logic of reality is interdependent on logic of cognition.
\cite{cite_obvious_strategyproofness_for_bounded_rationality_lead_to_good_approx} provides a trade-off between rationality of agents and approximation using obvious strategy proofness (OSP) in the context of machine scheduling problem. Authors argue that good approximations are possible if significant contingencies are taken into account within the bounded rationality constraints of the agents.
A restricted Turing machine in terms of running and storage has been used to model bounded rational agents to study repeated games is presented in \cite{cite_restricted_turing_machine_for_bounded_rationality}. This paper shows that when running time of the Turing machine is restricted it is harder to compute best response to a strategy than the strategy itself. If the storage space is restricted than best response to space restricted strategy cannot be implemented. Also, new Nash equilibria show up by restricting computational resources of agents in infinitely repeated games.
Authors in \cite{cite_human_decision_making_is_not_bayesian_rationality} argue that Bayesian rationality is not the way humans take decision rather they rely on heuristics to make decisions under certainties and risks. They survey recent developments on departures from Bayesian rationality and study bounded rationality, ecological rationality and nudges.
\cite{cite_bounded_rationality_relation_to_computation_and_ai} explains the detail relationship between bounded rationality and computational theories and argues that the latter can provide a better analytical explanation to bounded rationality.
In this work \cite{cite_bounded_rationality_to_understand_world_political_economy}, authors argues for sounder understanding of bounded rationality would improve the knowledge about world political economy empirically, theoretically, and practically.

\section{Discussion}
From the above research literatures, it is evident that there are diverse views on this topic of rationality with the advances in computing, AI, science and technology, etc. Majority of the research works believe that higher rationality levels (towards perfect rationality) is a possibility, though there are a few studies who tend to disagree. Also, there are several works which suggest that better understanding of bounded rationality is expected to happen. However, there is also some scepticism on judicious use of these advances for serving humanity in a better way. There are several models which are limited in scope and have to be broadened to understand the implications on mankind in a holistic sense.

The degrees and the dimensions of enhanced rationality that are going to impact mankind either way is an open question for further research. Whereas the hope is always for the best, but, time will show the consequences, and mankind will experience those changes whichever way these developments proceed in future.
\section{Conclusion}
Rationality has been a intriguing topic for a long time. The scope and the definition of rationality differ across subjects. A number of theories evolved from the concept of perfect rationality of agents (e.g. humans). However, the concept of perfect rationality among humans have been contradicted by the theory of bounded rationality. With the recent advances in computing, AI, science and technology, etc., there is an increasing interest on whether the shortcomings of bounded rationality can be augmented with machine intelligence leading humans to higher level of rationality.
This paper presents a recent survey (last five years) of some very interesting ideas and developments on this topic. Three schools of thoughts emerge from the survey - firstly, human bounded rationality and intelligence will not be exceeded by machines, secondly, limits of bounded rationality getting extended with augmented machine intelligence, and thirdly, better understanding of bounded rationality. However, there is scepticism on the ethical usage and consequences of these developments. Holistic study and understanding the consequences of enhanced rationality, augmented with machine intelligence, on mankind in the coming decades have to be further explored.
\bibliographystyle{ACM-Reference-Format}
\bibliography{rationality_survey}

\end{document}